\documentclass[11pt]{article}

\usepackage[a4paper,margin=1in]{geometry}
\usepackage{amsmath,amssymb}
\usepackage{graphicx}
\usepackage[round,authoryear]{natbib}
\setcitestyle{aysep={}}   % "(Bewley et al. 2001)", no comma before the year, as in JFM
\usepackage[font=small,labelfont=bf,labelsep=period]{caption}
\usepackage{authblk}
\usepackage{hyperref}
\hypersetup{colorlinks=true, urlcolor=blue, citecolor=black, linkcolor=black}

\numberwithin{equation}{section}

\title{Reaching separated flow states by release and recapture with learned latent feedback}

\author[1]{Khalid Rafiq}
\author[1]{Aditya G. Nair\thanks{Corresponding author: \href{mailto:adityan@unr.edu}{adityan@unr.edu}}}
\affil[1]{Department of Mechanical Engineering, University of Nevada, Reno, NV 89557, USA}
\date{}

\begin{document}

\maketitle

\begin{abstract}
Targeting a prescribed flow configuration rather than an aerodynamic coefficient is challenging because the desired state may lie outside the natural attractor. We demonstrate cluster-based control in a learned vorticity representation for two-dimensional flow over a NACA0015 airfoil at $Re=500$ and $\alpha=15^\circ$, using a localized streamwise momentum source near the leading edge. An actuation-enriched ensemble is compressed with a variational autoencoder and partitioned into macrostates. We optimize a continuous feedback law interpolating macrostate-dependent amplitudes without an explicit latent dynamical model. For an attached target, the optimized law approaches constant forcing, arrests vortex shedding and maintains the attached state. For a strongly separated target, the law instead establishes attachment, withdraws actuation so that the natural separation transient carries the flow through the target, and restores forcing to recapture attachment. Over the evaluation horizon, the closed-loop trajectory revisits the target approximately every three shedding periods. Companion open-loop calculations isolate the arrest, release and recapture processes.
\end{abstract}

\section{Introduction}
\label{sec:introduction}

Feedback control of separated and unsteady flows is usually formulated in terms of aerodynamic observables such as lift or drag \citep[e.g.][]{rabault2019artificial}. The response to actuation depends on the spatial organization of the flow. Similar instantaneous forces can correspond to different flow configurations, with different subsequent responses to actuation. Prescribing a flow configuration as the target therefore raises a question that aerodynamic objectives do not pose: which configurations can feedback reach and by what route?

Classical optimal-control formulations have addressed velocity-field tracking, full-state stabilization, and trajectory tracking \citep{gunzburger2000tracking,bewley2001dns,hogberg2003relaminarization}. Repeated solutions of the governing and adjoint equations are computationally demanding \citep{bewley2001dns}. Feedback has also been used to stabilize nonlinear invariant states extracted from turbulent trajectories \citep{yasuda2025travelling}. Reduced descriptions capture relevant dynamics in a few coordinates \citep{rowley2017model}. Linearized and resolvent-based models are particularly effective for dynamics organized around a base or mean flow \citep{yeh2019resolvent}. Steering between attached and separated configurations, however, may involve nonlinear excursions beyond the natural attractor. The reduced coordinates must therefore distinguish these configurations and the transient pathways connecting them.

Cluster-based descriptions coarse-grain nonlinear trajectories into representative states \citep{kaiser2014cluster,li2021cluster} and have been used to control separating flows \citep{kaiser2017cluster}. \citet{nair2019cluster} clustered aerodynamic-force trajectories of a post-stall airfoil and optimized cluster-dependent actuation to redirect the dynamics toward favorable regions of a low-dimensional feature space. Nonlinear encoders extract compact coordinates directly from flow fields and have been used for reduced-order modeling and control \citep{morton2018deep,ishize2024flow}, including learned embeddings of separated flows \citep{fukami2023grasping,tran2026optimal} and latent-space cluster control for the suppression of extreme events \citep{shehata2026data}. Here, we combine a learned flow-field representation with cluster-based control to target prescribed full-field configurations. We learn the representation from an actuation-enriched ensemble that includes both natural and forced flow configurations. We then cluster the encoded trajectories and optimize cluster-dependent actuation amplitudes against a target-based objective, without constructing an explicit latent dynamical model.

We demonstrate the approach for two-dimensional flow over a NACA0015 airfoil at $Re=500$ and $\alpha=15^\circ$ with a localized, non-negative streamwise momentum source near the leading edge. Because momentum addition promotes attachment, the two targets present different control problems: forcing can maintain an attached state but does not directly sustain a strongly separated one. Changes in actuation initiate finite-time reattachment and separation through the formation and convection of large-scale vortical structures \citep{amitay2002controlled,darabi2004part1,darabi2004part2,siauw2017transient}, and these transients prove central to the control. For the attached target, the optimized feedback arrests vortex shedding and holds the flow near the target. For the separated target, it produces a release-and-recapture cycle: it first drives the flow to attachment, withdraws actuation so that the separation transient passes through the target, and then restores forcing to recapture the attached state. The separation transient thus provides a route to the target where withdrawing actuation becomes an important control action.

\section{Methodology}
\label{sec:method}

\subsection{Flow configuration, actuation and data ensemble}
\label{sec:flow}

\begin{figure}
\centerline{\includegraphics[width=0.9\linewidth]{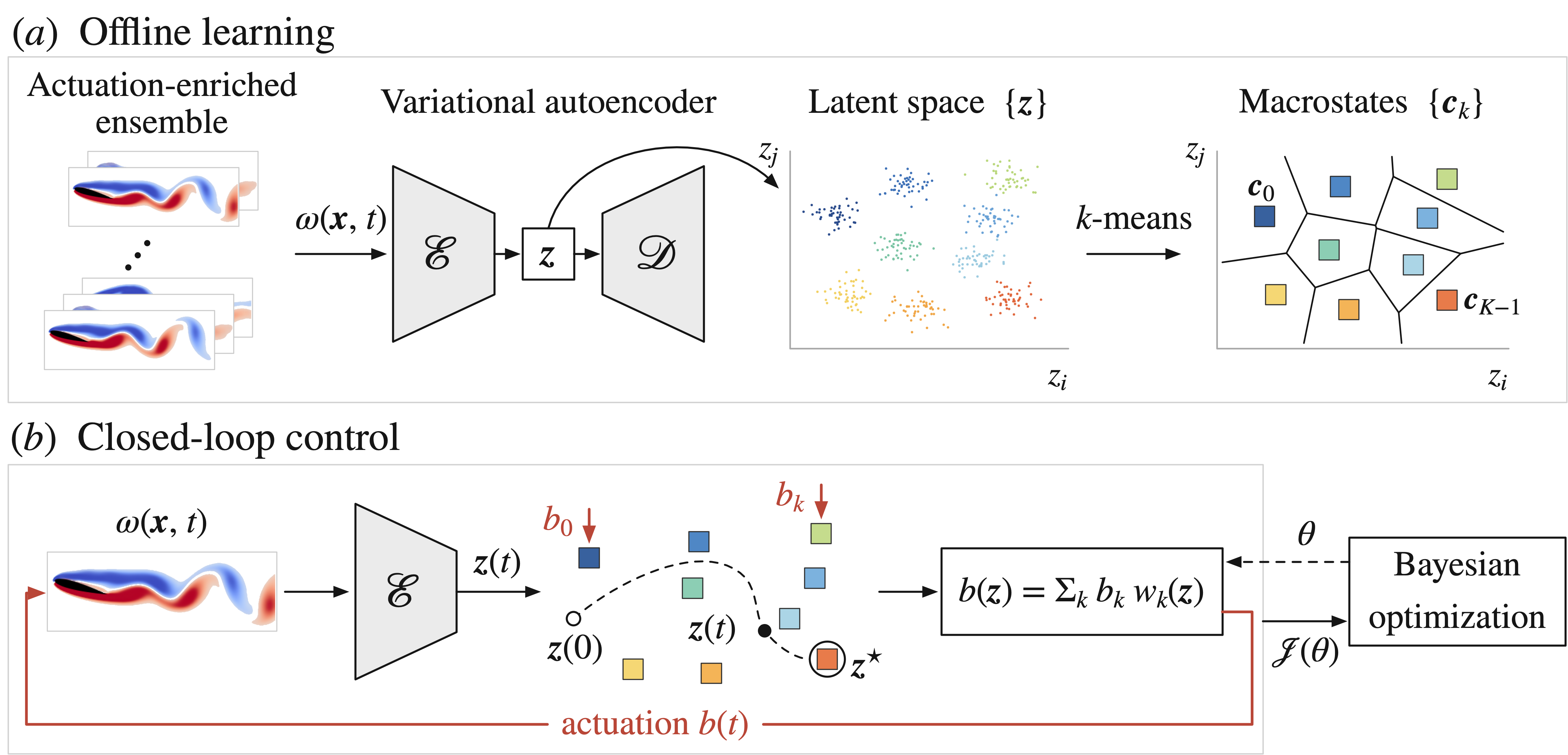}}
\caption{Overview of the approach. (\textit{a}) An actuation-enriched ensemble of vorticity fields is encoded by a variational autoencoder into $\boldsymbol z\in\mathbb R^d$ and partitioned by $k$-means into $K$ macrostates. (\textit{b}) In closed loop, the encoded state sets the forcing through the cluster amplitudes $b_k$ and kernel width $\sigma_w$ in (\ref{eq:RBF}), with $w_k$ the normalized Gaussian weight of cluster $k$. Each episode evaluates one parameter set $\theta$, and Bayesian optimization minimizes $\mathcal J(\theta)$ in (\ref{eq:cost}); $\boldsymbol z^\star$ is the target.}
\label{fig:overview}
\end{figure}

We consider two-dimensional flow over a NACA0015 airfoil at $Re=U_\infty c/\nu=500$ and $\alpha=15^\circ$, with lengths, velocities and times nondimensionalized by the chord $c$, the freestream velocity $U_\infty$ and $c/U_\infty$ ($\nu$ is the kinematic viscosity). The incompressible Navier--Stokes equations,
\begin{equation}
\frac{\partial \boldsymbol q}{\partial t}+\boldsymbol q\boldsymbol{\cdot}\nabla\boldsymbol q
=-\nabla p+\frac{1}{Re}\nabla^2\boldsymbol q+\boldsymbol f(\boldsymbol x,t),
\qquad \nabla\boldsymbol{\cdot}\boldsymbol q=0,
\label{eq:NS}
\end{equation}
where $\boldsymbol q$, $p$ and $\boldsymbol f$ are the velocity, pressure and external forcing, are solved with a multi-domain immersed-boundary projection method \citep{colonius2008fast,goza2017strongly} on five nested $288\times288$ grids. The innermost domain spans $(x/c,y/c)\in[-1,1]^2$ with $\Delta x/c\simeq0.0069$, the outermost domain extends to 16 chords and the time step is $\Delta t=0.001$. The unforced flow exhibits periodic vortex shedding with mean drag coefficient $\overline C_D=0.29$ and dominant frequency $f_{\rm sh}\simeq0.58$, i.e. a shedding period $T_{\rm sh}\simeq1.73$. Refinement to nested $360\times360$ grids with similar domain extents changes $f_{\rm sh}$, $\overline C_L$ and $\overline C_D$ by less than $0.5\%$, $1.3\%$ and $0.3\%$, respectively.

Control is applied through a localized momentum source,
\begin{equation}
\boldsymbol f(\boldsymbol x,t)=b(t)\,\delta_\eta(\boldsymbol x-\boldsymbol x_f)\,\hat{\boldsymbol e}_f,
\label{eq:forcing}
\end{equation}
where $b(t)$ is the scalar forcing amplitude, $\boldsymbol x_f$ and $\hat{\boldsymbol e}_f$ denote the fixed actuator location and direction, and $\delta_\eta$ is a compact regularized approximation of the Dirac delta with support half-width $\eta=3\Delta x$. The actuator is centered $3\Delta x$ above the suction surface at $7.5\%$ chord, $(x_f/c,y_f/c)\simeq(-0.155,0.117)$, near leading-edge separation, and acts in the freestream ($x$) direction. We characterize actuation by $C_\mu=\rho_\infty u_{\rm jet}^2\sigma/(\tfrac12\rho_\infty U_\infty^2c)$, where $\rho_\infty$ is the freestream density, $\sigma=\Delta x$ and $u_{\rm jet}$ is the peak time-averaged velocity in a companion simulation without free stream \citep{munday2013lock}. For $b=0.05$, $u_{\rm jet}/U_\infty=1.79$ and $C_\mu\approx0.044$. We constrain $b\geq0$, so the controller can only add streamwise momentum.

An actuation-enriched ensemble is generated with open-loop forcing $b(t)=\tfrac12A[1+\sin(2\pi f_a t)]$ for four amplitudes $A\in\{0.02,0.04,0.06,0.08\}$ and 14 frequencies $f_a/f_{\rm sh}\in\{0.1,0.25,0.5,0.75,1,\ldots,3.25\}$, giving 56 forcing conditions. Vorticity snapshots are stored every $50\Delta t$ over the statistically developed window $t\in[80,100]$, 400 per case and $M=22\,400$ in total. We use 44 forcing conditions for training and 12 held-out conditions for validation.

Figure~\ref{fig:overview} summarizes the approach (\S\ref{sec:latent}--\ref{sec:control}). The encoder and centroids remain fixed during optimization.

\subsection{Latent representation}
\label{sec:latent}

The encoder takes the near-body vorticity field $\Omega$ over $x/c\in[-0.4,\,3.8]$ and $y/c\in[-0.85,\,0.85]$. We render this field on a uniform $128\times320$ grid as a three-channel image with a fixed color scale saturating at $|\Omega c/U_\infty|=4$. This saturation suppresses near-wall vorticity by design, emphasizing shear-layer and wake topology rather than boundary-layer detail. We denote the resulting input by $\omega$. A variational autoencoder (VAE) \citep{kingma2013auto} maps $\omega$ to a Gaussian latent distribution $q(\widetilde{\boldsymbol z}\mid\omega)=\mathcal N\big(\boldsymbol\mu(\omega),\operatorname{diag}\boldsymbol\sigma^2(\omega)\big)$. During training, the latent variable is sampled as $\widetilde{\boldsymbol z}=\boldsymbol\mu(\omega)+\boldsymbol\sigma(\omega)\odot\boldsymbol\epsilon$, with $\boldsymbol\epsilon\sim\mathcal N(\boldsymbol 0,\boldsymbol I)$ and $\widetilde{\boldsymbol z}\in\mathbb R^d$, and the decoder reconstructs the field as $\widehat\omega=\mathcal D(\widetilde{\boldsymbol z})$ minimizing
\begin{equation}
\mathcal L=\frac{1}{N}\,\mathbb E\!\left[\|\omega-\mathcal D(\widetilde{\boldsymbol z})\|_2^2\right]
+\frac{\beta}{d}\,D_{\rm KL}\!\left(q(\widetilde{\boldsymbol z}\mid\omega)\,\|\,\mathcal N(\boldsymbol 0,\boldsymbol I)\right),
\qquad \beta=5\times10^{-3},
\label{eq:vae}
\end{equation}
where $N$ is the number of values in $\omega$ and $d$ the latent dimension, so that the reconstruction and Kullback--Leibler terms are both per-element averages. Training uses AdamW for 250 epochs (learning rate $3\times10^{-4}$ with cosine decay to $10^{-6}$, weight decay $10^{-4}$, batch size 128, gradient-norm clipping at unity), retaining the weights with the lowest validation loss. For clustering and feedback, we use the posterior mean $\boldsymbol z(t)=\boldsymbol\mu(\omega(t))$.

We compare plain and residual convolutional VAEs (the latter with two-layer refinement blocks of $3\times3$ convolutions, GroupNorm and GELU activations) for $d\in\{2,3,4,6,8,16,32,64\}$. The validation reconstruction error decreases with latent dimension [figure~\ref{fig:latent_diagnostics}$(a)$], and we select $d=8$ as a compromise between reconstruction accuracy and a compact coordinate system for clustering and feedback. At this dimension, the residual VAE ($44.4$ million parameters) gives a lower validation error than the plain VAE ($42.5$ million) and an error comparable to POD with $r=61$ modes. 

Euclidean latent distance defines cluster membership, feedback weights and target proximity. It is not assumed to be a physical norm between flow fields without further constraints on the learned geometry \citep{tran2026optimal}. We assess the physical significance of the macrostates through their flow fields, loads and trajectory dynamics.

\subsection{Cluster-based latent dynamics}
\label{sec:clustering}

\begin{figure}
\centerline{\includegraphics[width=0.92\linewidth]{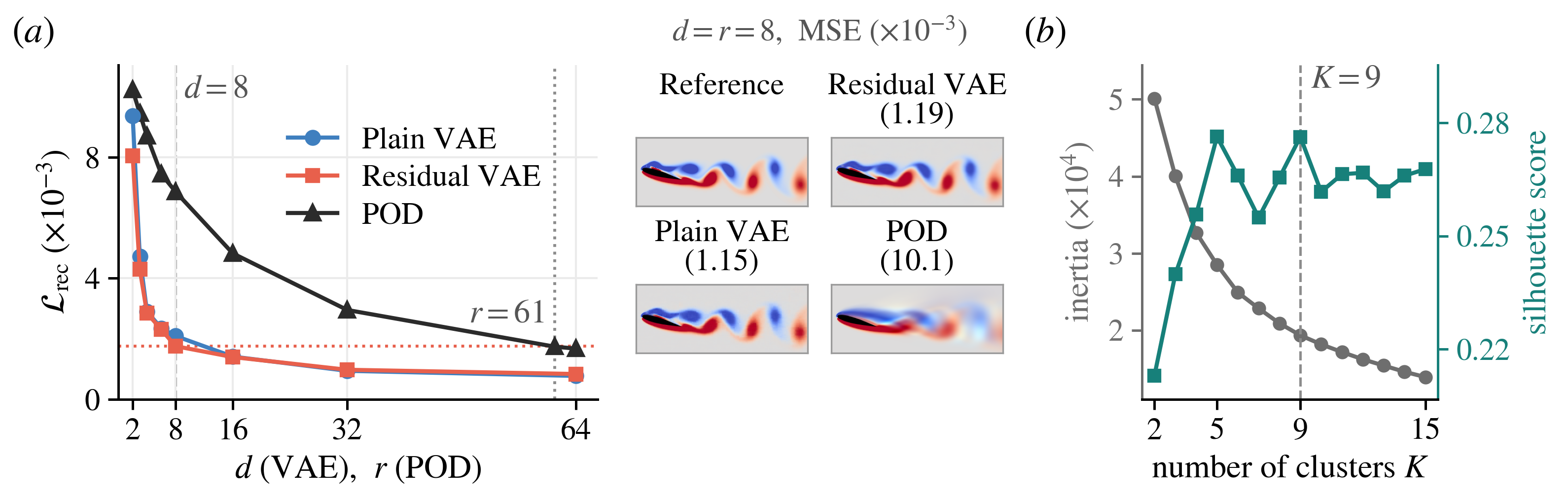}}
\caption{Latent representation and clustering diagnostics. $(a)$ Validation reconstruction error $\mathcal L_{\rm rec}$ (per-element mean-squared error) of the plain and residual VAEs, decoded from the posterior mean, versus dimension $d$, and of POD versus number of modes $r$. Insets show a validation snapshot and reconstructions at $d=r=8$; parentheses give snapshot MSE ($\times10^{-3}$). $(b)$ Within-cluster inertia (left axis) and silhouette score (right axis) versus $K$ at $d=8$; the dashed line marks $K=9$.}
\label{fig:latent_diagnostics}
\end{figure}

The encoded open-loop trajectories are partitioned by $k$-means. For each $K=2$--15, $k$-means++ is initialized 100 times and the minimum-inertia solution is retained. Within-cluster inertia decreases with $K$, while the silhouette score (computed on a random subsample of 8000 snapshots) exhibits comparable peaks at $K=5$ and $K=9$ [figure~\ref{fig:latent_diagnostics}$(b)$]. We select $K=9$ for a finer partition without substantially reducing the silhouette score. The maximum score, about $0.28$, indicates that the clusters divide continuous trajectories rather than isolated regimes. We describe their dynamics by transition probabilities and residence times (individual visit durations). For each cluster, we record the probability of the next cluster visited, the mean visit duration $\bar\tau_k$, and the dwell time conditioned on the cluster from which it was entered. Times are normalized by $T_{\rm sh}$.

\subsection{Cluster-based feedback and optimization}
\label{sec:control}

Following \citet{nair2019cluster}, the cluster amplitudes $b_k$ are interpolated to define the feedback law
\begin{equation}
b(\boldsymbol z)=\frac{\sum\nolimits_{k=0}^{K-1} b_k\,e^{-\|\boldsymbol z-\boldsymbol c_k\|_2^2/(2\sigma_w^2)}}
{\sum\nolimits_{k=0}^{K-1} e^{-\|\boldsymbol z-\boldsymbol c_k\|_2^2/(2\sigma_w^2)}},
\label{eq:RBF}
\end{equation}
where $\boldsymbol c_k$ is the centroid of $C_k$ and $\sigma_w>0$ is the interpolation width. Thus, the forcing in (\ref{eq:forcing}) is a smooth, distance-weighted average of the cluster amplitudes. The loop is closed by encoding the instantaneous flow field, $\boldsymbol z(t)=\boldsymbol\mu(\omega(t))$. For a prescribed target $\boldsymbol z^\star$, the parameters $\theta=\{b_0,\ldots,b_{K-1},\sigma_w\}$ are selected to minimize
\begin{equation}
\mathcal J(\theta)=w_{\min}\min_t\|\boldsymbol z(t;\theta)-\boldsymbol z^\star\|_2
+w_{\rm mean}\left\langle\|\boldsymbol z(t;\theta)-\boldsymbol z^\star\|_2\right\rangle_t
+w_E\left\langle b(t)^2\right\rangle_t.
\label{eq:cost}
\end{equation}
The terms reward the closest approach, measure mean target distance over the evaluation horizon, and penalize actuation effort, respectively. We use $(w_{\min},w_{\rm mean},w_E)=(0.6,0.2,30)$. The effort weight is larger because $\langle b^2\rangle$ ($\sim10^{-3}$ for the reported laws) is much smaller than latent distances of order unity. The effort term thus contributes 4--12\% of $\mathcal J$ for the escape laws and 18--35\% for the hold laws. The target is the realized latent state nearest the selected centroid, ensuring it corresponds to an actual flow realization.

All candidate laws start from the same baseline realization and are evaluated for $20\,000$ solver steps, approximately $12T_{\rm sh}$. The flow field is encoded and the forcing updated every ten steps, then held constant. The distance terms in (\ref{eq:cost}) are evaluated every 50 solver steps over the whole episode, including the initial transient, and the effort term over all forcing updates. The parameters are determined by Bayesian optimization (Weights \& Biases implementation; Gaussian-process surrogate with a Mat\'ern kernel and expected improvement) with search bounds $b_k\in[0,0.10]$ and $\sigma_w\in[0.25,2.0]$. The searches for the attached and separated targets evaluate approximately 40 and 80 candidate laws. Given these budgets in ten dimensions, the reported laws are the best found rather than verified global optima. We examine the ten lowest-loss laws from each search to assess their structural consistency.

\section{Results}
\label{sec:results}

\subsection{Macrostates of the actuated flow}
\label{sec:states}

\begin{figure}
\centerline{\includegraphics[width=0.9\linewidth]{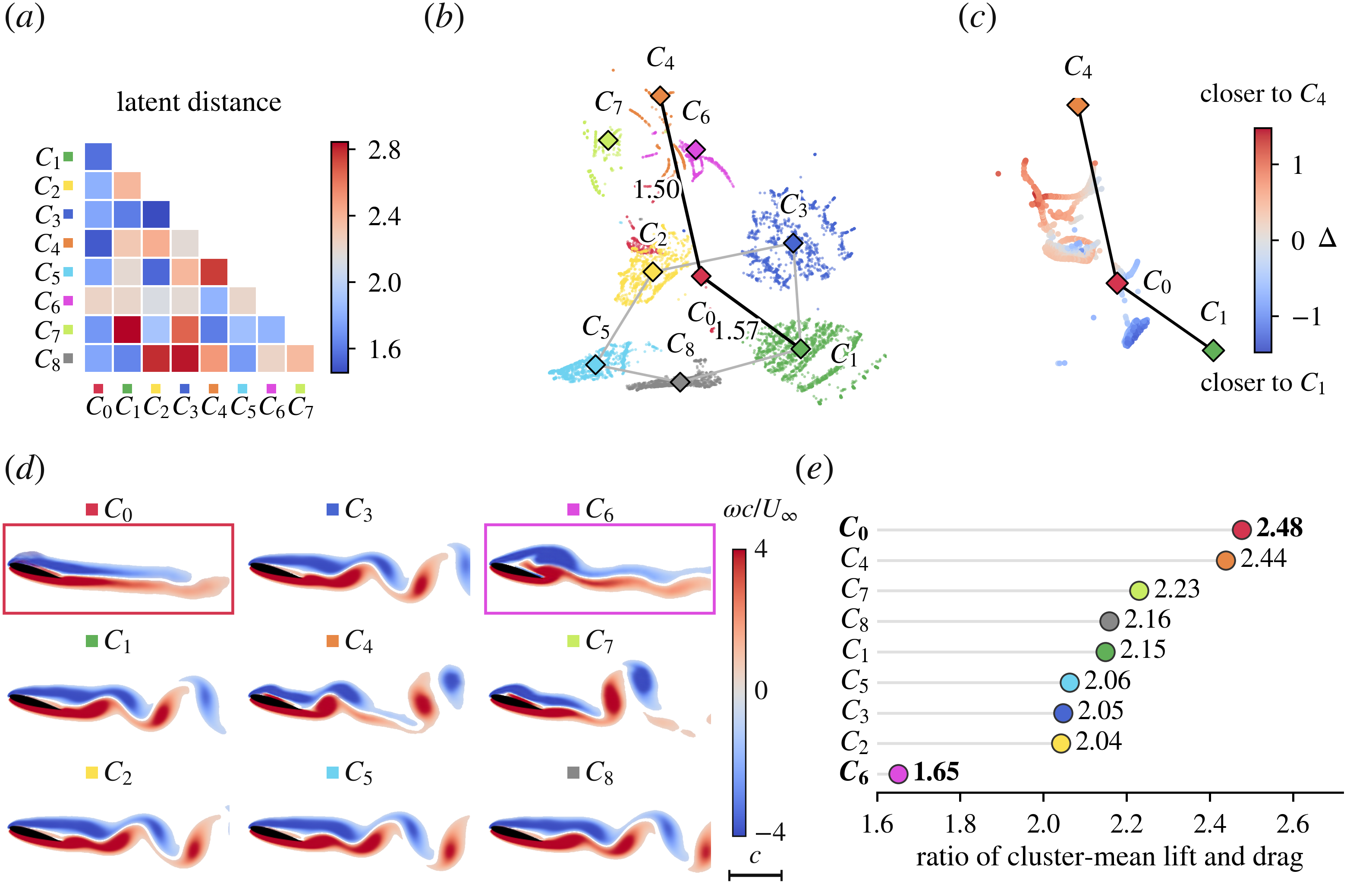}}
\caption{Macrostates of the actuated flow. $(a)$ Pairwise centroid distances in the eight-dimensional latent space. $(b)$ UMAP visualization of the clustered trajectories; the gray ring links the shedding states in their baseline order, and the annotated links report distances evaluated in the original latent coordinates. $(c)$ Distribution of $C_0$ relative to $C_1$ and $C_4$, colored by $\Delta=\|\boldsymbol z-\boldsymbol c_1\|_2-\|\boldsymbol z-\boldsymbol c_4\|_2$. $(d)$ Decoded centroid fields on a common color scale, with outlines marking the target macrostates $C_0$ and $C_6$. $(e)$ Ratio of cluster-mean lift and drag coefficients, $\langle C_L\rangle_{C_k}/\langle C_D\rangle_{C_k}$.}
\label{fig:states}
\end{figure}

Figure~\ref{fig:states} summarizes the nine macrostates. The decoded centroid fields [figure~\ref{fig:states}$(d)$] show that $C_1,C_2,C_3,C_5$ and $C_8$ represent phases of vortex shedding, while $C_0$ represents an attached configuration with a thin, elongated wake. The states $C_4,C_6$ and $C_7$ occur during separation, including the roll-up and convection of a large vortical structure over the suction surface.

In the eight-dimensional latent space, $C_0$ lies between the shedding and separated branches. It is closest to $C_4$ and $C_1$ (distances $1.50$ and $1.57$), which are themselves farther apart [figure~\ref{fig:states}$(a,b)$]. The UMAP visualization \citep{mcinnes2018umap} shows these trajectories [figure~\ref{fig:states}$(b)$], while figure~\ref{fig:states}$(c)$ shows that some members of $C_0$ lie closer to $C_1$ and others to $C_4$. This position becomes important under control.

The ratio of cluster-mean lift and drag coefficients [figure~\ref{fig:states}$(e)$] is highest for $C_0$, $\langle C_L\rangle_{C_0}/\langle C_D\rangle_{C_0}=2.48$, and lowest for $C_6$, $1.65$. We therefore select $C_0$ and $C_6$ as contrasting attached and separated targets, not as aerodynamic performance objectives. We refer to the laws targeting them as hold and escape laws, respectively. Notably, $C_4$ has a ratio of $2.44$, close to that of $C_0$, despite a markedly different flow structure. Thus, this aerodynamic measure alone cannot distinguish the two configurations, motivating the field-based representation.

\subsection{Optimized control of the coarse-grained dynamics}
\label{sec:controlled}

\begin{figure}
\centerline{\includegraphics[width=0.92\linewidth]{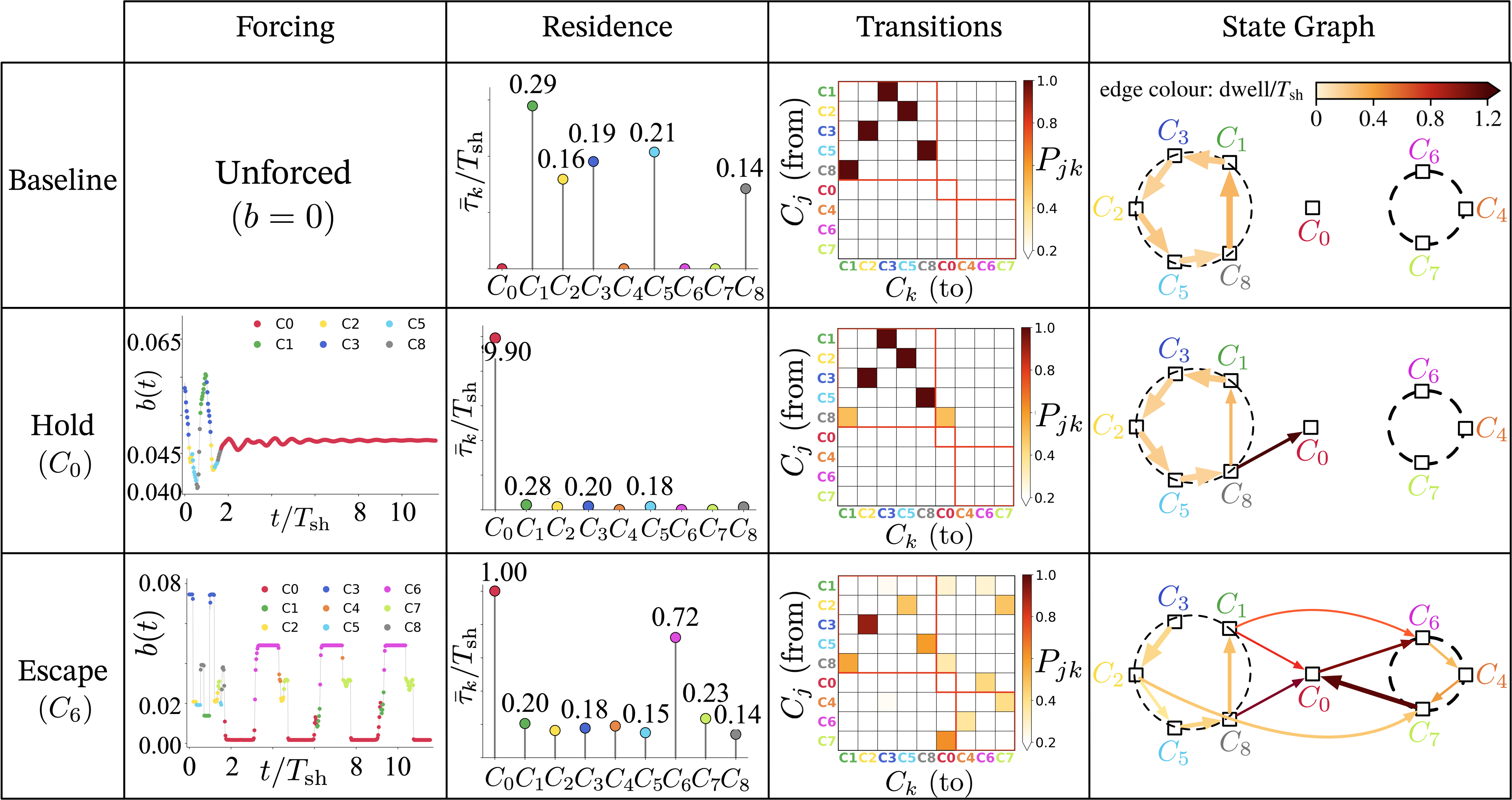}}
\caption{Coarse-grained dynamics for the unforced baseline (top), hold laws targeting $C_0$ (middle) and escape laws targeting $C_6$ (bottom). Columns show representative forcing colored by macrostate; mean observed residence time per visit; transition probability $P_{jk}$ from $C_j$ (row) to $C_k$ (column), with cells below $0.2$ white; and a state graph of transitions observed at least ten times with conditional probability above $0.2$. Graph edge width denotes transition probability and edge color the conditional destination dwell time (saturating at $1.2T_{\rm sh}$). Dashed circles group the shedding and separated states.}
\label{fig:control}
\end{figure}

Figure~\ref{fig:control} compares the unforced baseline (top row), hold laws targeting $C_0$ (middle row) and escape laws targeting $C_6$ (bottom row). The forcing histories are representative trajectories; the controlled residence-time and transition statistics aggregate the ten lowest-loss laws for each target. The unforced flow follows the shedding sequence $C_1\rightarrow C_3\rightarrow C_2\rightarrow C_5\rightarrow C_8\rightarrow C_1$, with residence times from $0.14T_{\rm sh}$ in $C_8$ to $0.29T_{\rm sh}$ in $C_1$. Neither $C_0$ nor the separated branch is visited during the baseline episode.

Under the hold laws, the representative trajectory enters $C_0$ after approximately $1.7T_{\rm sh}$. Forcing settles near $b\simeq0.047$ with small residual variations, and the trajectory remains in $C_0$ for the episode. Across the ten laws, the mean observed visit duration in $C_0$ is $9.90T_{\rm sh}$, compared with less than $0.30T_{\rm sh}$ in each visited shedding state. Visits interrupted by the start or end of an episode are counted at their observed length, so the residence in $C_0$ is bounded by the evaluation horizon. The transition matrix and state graph show access to $C_0$ from $C_8$.

Under the escape laws, forcing of up to $b\simeq0.08$ during the initial shedding transient brings the trajectory to $C_0$. The forcing then falls to nearly zero, allowing the flow to enter the separated target $C_6$. There, forcing rises to $b\simeq0.05$, and the trajectory passes through intermediate states before returning to $C_0$. In the representative history, this sequence repeats, with visits to $C_6$ approximately every $3T_{\rm sh}$ ($2.8T_{\rm sh}$ on average over the ten lowest-loss laws, range $2.2$--$3.4T_{\rm sh}$). Thus, the feedback repeatedly reaches the target but does not hold the flow there. The mean visit durations in $C_0$ and $C_6$ are $T_{\rm sh}$ and $0.72T_{\rm sh}$, respectively. 
Across the ten laws, the transition matrix and state graph show the pathway $C_0\rightarrow C_6\rightarrow C_4\rightarrow C_7\rightarrow C_0$. They also show direct entries into $C_6$ from shedding states (for example, $C_1\rightarrow C_6$); release from $C_0$ is therefore not the only route to the target. In the representative history, however, each visit to $C_6$ follows the withdrawal of forcing in $C_0$ and is followed by renewed forcing, forming a release--recapture cycle.

The ten lowest-loss escape laws assign little forcing to $C_0$ but positive forcing to $C_6$. In every case, $b_0\leq0.0052$ (median $0.002$), whereas $b_6\geq0.022$ (median $0.049$). The hold laws, by comparison, retain a median $b_0=0.034$ under the same effort penalty. This contrast indicates that the near-zero $b_0$ in the escape laws reflects the target objective rather than the penalty alone.

The escape laws also use narrow interpolation kernels, with median
$\sigma_w=0.298$ (range $0.259$--$0.453$), much smaller than the centroid
spacing of $1.5$--$2.8$ [figure~\ref{fig:states}$(a)$]. The forcing
therefore approaches piecewise-constant behavior, with individual
macrostate amplitudes dominating and sharp changes as the trajectory
moves between macrostates. In contrast, the hold laws have median
$\sigma_w=1.514$ (range $0.540$--$1.700$), comparable to the centroid
spacing. This blends neighboring amplitudes and produces smoother
forcing, varying between $0.040$ and $0.065$ during the initial
shedding transient.

The remaining macrostate amplitudes vary more across the ten laws,
especially for the hold objective. Because the applied forcing is a
kernel-weighted blend, amplitudes associated with rarely visited or
strongly blended macrostates are only weakly constrained by the cost.
These amplitudes can differ without appreciably changing the applied
forcing. Together with the observed forcing histories, these results
suggest a three-stage state-dependent mechanism: forcing promotes
attachment (arrest), its withdrawal permits a separation transient
(release), and renewed forcing directs the flow back toward $C_0$
(recapture).

\subsection{Arrest, release and recapture}
\label{sec:mechanism}

We use open-loop calculations to examine the three processes underlying the optimized escape laws: maintaining attachment, allowing separation, and restoring attachment. By controlling when forcing is applied or removed, we examine each process separately.

\begin{figure}
\centerline{\includegraphics[width=0.92\linewidth]{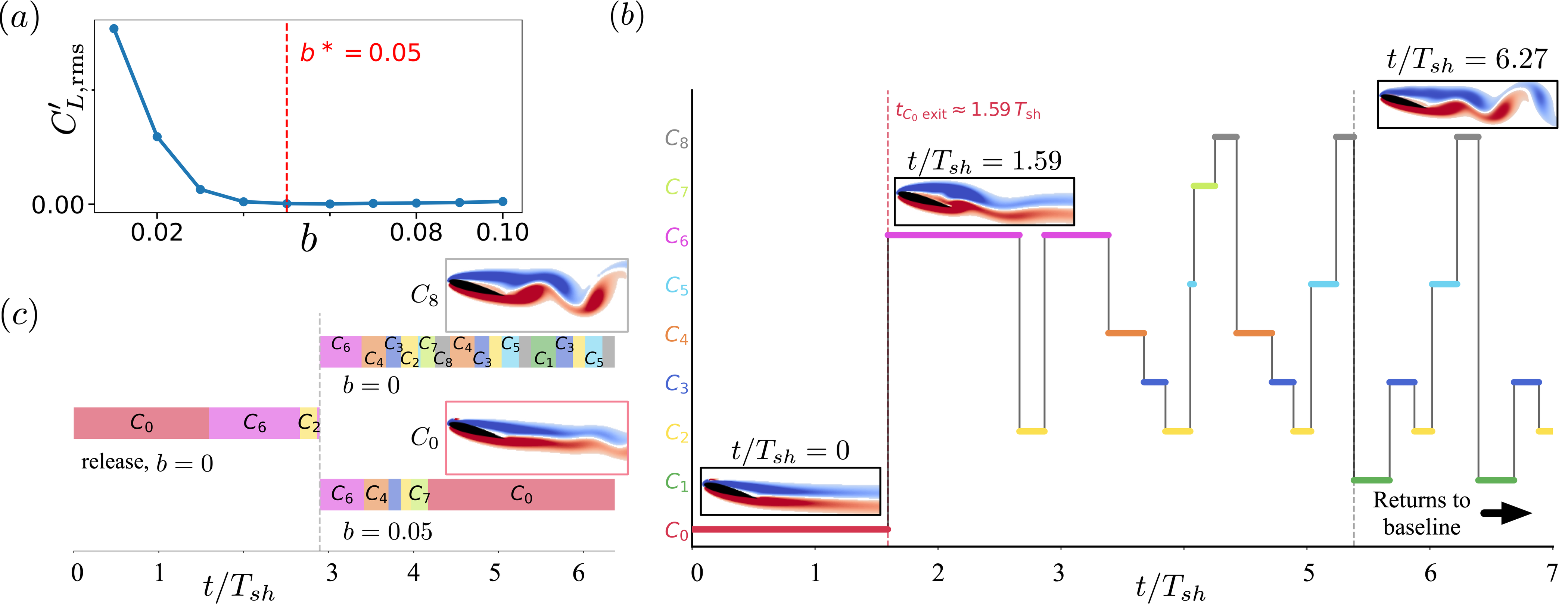}}
\caption{Open-loop calculations isolating arrest, release and recapture. $(a)$ Constant-forcing sweep initialized from settled $C_0$; residual lift oscillations become negligible by $b^\star=0.05$. $(b)$ Unforced release from $C_0$, showing entry into $C_6$ after approximately $1.59T_{\rm sh}$ and subsequent return to the shedding states. $(c)$ Trajectories sharing the same release history until $t/T_{\rm sh}=2.89$, after which the upper branch remains unforced and the lower branch receives $b=b^\star$. Timeline colors denote the instantaneous macrostate; the re-actuated branch is extended to show continued residence in $C_0$.}
\label{fig:probe}
\end{figure}

\textit{Arrest.} We first determine the forcing needed to maintain an attached flow. Starting from a settled $C_0$ state (the final flow field of a hold-law episode), we apply constant amplitudes $b\in[0.01,0.10]$ for $12T_{\rm sh}$. We measure the residual lift fluctuations, $C'_{L,{\rm rms}}=\langle(C_L-\overline{C_L})^2\rangle^{1/2}$, over the final five shedding periods. Figure~\ref{fig:probe}$(a)$ shows that these fluctuations decrease rapidly with increasing forcing and become negligible by $b^\star=0.05$. This value is close to the settled hold-law forcing ($b\simeq0.047$) and the median escape-law amplitude assigned to $C_6$ ($b_6=0.049$). The sweep therefore provides a reference forcing level for maintaining attachment, although it does not establish the forcing needed to reach attachment from the shedding cycle.

\textit{Release.} We next examine what happens when forcing is removed from the settled $C_0$ state. The flow remains in $C_0$ for approximately $1.59T_{\rm sh}$ before entering the separated target $C_6$ [figure~\ref{fig:probe}$(b)$]. A large vortical structure then develops and convects over the suction surface, and the flow eventually returns to periodic shedding at $t/T_{\rm sh}\approx5.4$. Thus, the natural separation transient passes through $C_6$, providing a route to the target simply by withdrawing forcing.

Under the escape laws, the mean residence time in $C_0$ is shorter, approximately $T_{\rm sh}$. This difference is consistent with the different conditions: the open-loop calculation starts from fully settled $C_0$, whereas the controlled trajectory re-enters $C_0$ after recapture and experiences little forcing there (median assigned amplitude $b_0=0.002$). The two residence times therefore describe different initial conditions.

\textit{Recapture.} Finally, we examine whether restoring forcing during separation can return the flow to attachment. Figure~\ref{fig:probe}$(c)$ compares two trajectories that undergo the same unforced release until $t/T_{\rm sh}=2.89$. One remains unforced, while the other receives constant forcing $b=b^\star=0.05$. Both initially follow similar paths, indicating that renewed forcing does not immediately reverse the separation transient. Their subsequent evolution, however, differs. The unforced flow returns to periodic shedding near $t/T_{\rm sh}\approx5.4$, whereas the re-actuated flow passes through $C_7$ and reaches the attached state $C_0$ at $t/T_{\rm sh}\simeq4.17$, approximately $1.3T_{\rm sh}$ after forcing resumes. Restoring forcing during separation therefore recaptures attachment before the flow returns to its natural shedding cycle.

This result explains the role of positive $b_6$ in the optimized escape laws. Forcing in $C_6$ does not maintain the separated target; it redirects the flow toward $C_0$, from which another release can occur. The combination of near-zero $b_0$ and positive $b_6$ thus allows the feedback to alternate between releasing the flow toward the target and recapturing attachment. The short mean residence time in $C_6$ under feedback ($0.72T_{\rm sh}$) is consistent with this mechanism, although the open-loop and controlled trajectories enter $C_6$ under different conditions. These finite-time responses are consistent with observations of controlled separation and reattachment over stalled airfoils \citep{amitay2002controlled,darabi2004part1,darabi2004part2,siauw2017transient}. Here, however, the arrest--release--recapture sequence emerges from optimization rather than being prescribed.

\section{Concluding remarks}
\label{sec:conclusions}

We developed cluster-based feedback to target prescribed flow configurations of a stalled airfoil. The strategy optimizes macrostate-dependent actuation amplitudes in a learned eight-dimensional latent space, subject to a non-negative momentum-addition constraint. For the attached target $C_0$, the optimized law approaches constant forcing, at a level where residual lift oscillations become negligible, and maintains attachment over the evaluation horizon. For the separated target $C_6$, the law instead establishes attachment, withdraws forcing to let the separation transient pass through $C_6$, and then restores forcing to recapture $C_0$. Companion open-loop calculations isolate these three stages.

The main physical result is that target-state feedback can exploit, rather than oppose, the natural separation transient. Because the actuator can only add momentum, the flow reaches the separated target by withdrawing forcing and allowing separation to develop. The optimized law then restores forcing during the transient and recaptures $C_0$ before the flow returns to shedding. The objective selects this sequence without prescribing it, rewarding repeated approaches to the target at limited effort. Directing a nonlinear system to a target by exploiting its own dynamics is a classical idea in nonlinear control \citep{shinbrot1990using}. Here a related strategy emerges from optimization, for a separated flow, in a learned coordinate system.

The analysis is limited to a two-dimensional, low-Reynolds-number flow, one actuator and a fixed initial shedding phase. Target proximity is evaluated in learned coordinates, and feedback requires the full vorticity field. Within this scope, the representation identifies transient macrostates, the flow and actuator determine accessible pathways, and optimization selects among them. Experimental implementation would require estimating latent coordinates from sparse measurements. Deep-learning estimators mapping sensor data to a reduced state \citep{nair2020leveraging} offer one route. Because forcing varies little within a macrostate, a coarse estimate of the latent state may suffice away from macrostate transitions.

\paragraph*{Funding.}
A.G.N.\ acknowledges support from the National Science Foundation AI Institute in Dynamic Systems (Award No.\ 2112085; 2624398), the Department of Energy Early Career Research Award (Award No.\ DE-SC0022945) and Air Force Office of Scientific Research (Award No.~FA9550-23-1-0483).

\paragraph*{Declaration of interests.}
The authors report no conflict of interest.

\paragraph*{Author ORCIDs.}
Aditya G. Nair, https://orcid.org/0000-0002-8979-8420.

\paragraph*{Data availability statement.}
The code, cluster centroids and optimized control-law parameters are available at \url{https://github.com/Khalid-Rafiq-01/cluster_based_latent_flow_control}; the trained encoder and simulation data are available from the corresponding author upon reasonable request.

\bibliographystyle{jfm}
\bibliography{references_updated}

\end{document}